\documentclass{appolb}
\usepackage{graphicx}
\usepackage{amsmath}
\usepackage{amssymb}
\usepackage{dsfont}

\begin{document}
\title{A comment on the generalization of the Marinatto-Weber quantum game scheme}
\author{Piotr Fr\c{a}ckiewicz
\address{Institute of Mathematics, Polish Academy of Sciences,\\ ul. \'Sniadeckich 8, 00-956 Warsaw, Poland}} \maketitle
\begin{abstract}
Iqbal and Toor [Phys. Rev. A {\bf 65}, 022306 (2002)] and [Commun.
Theor. Phys. {\bf 42}, 335 (2004)] generalized the Marinatto-Weber
quantum scheme for $2 \times 2$ games in order to study bimatrix
games of $3 \times 3$ dimension, in particular the
Rock-Paper-Scissors game. In our paper we show that Iqbal and
Toor's generalization exhibits certain undesirable property that
can considerably influence the game result. To support our
argumentation, in the further part of the paper we construct the
protocol corresponding to the MW concept for any finite bimatrix
game that is free from the fault.
\end{abstract}

\section{Introduction}
The Marinatto-Weber (MW) scheme \cite{mw} has become one of the
most frequently used quantum game schemes. Though it was created
for research on Nash equilibria in quantum $2\times 2$ games, it
has also found application in studying some of the refinements of
a Nash equilibrium like evolutionarily stable strategies
\cite{iqbal}. Moreover, it has been proved that the MW scheme is
applicable to finite extensive games \cite{fr} and even various
problems of duopoly \cite{iqbal2, krk, krk1}. These recent papers
show uninterrupted interest in research on quantum games played
according to the MW idea and they provide sufficient motivation to
improve the existing results.\\ \noindent Our comment mainly
concerns frequently cited paper \cite{iqbalfail}, where the
concept of quantum $3 \times 3$ game was introduced. However, it
also relates to the later paper \cite{iqbalfail2}, where the same
concept was used.

\section{Comment on the Iqbal and Toor's quantum $\mathbf{3 \times 3}$ game}

The possibility of recovering classical game from its quantum
counterpart is a necessary condition for any quantum game scheme
to be treated as a correct one. In the case of the MW scheme the
classical game is obtained by putting the initial state
$|\psi_{\mathrm{in}}\rangle = |00\rangle$. In fact, the
Marinatto-Weber construction allows us to recover the classical
game by putting any basis state $|ij\rangle$, $i,j = 0,1$ as the
initial state. If, for example, the initial state equals
$|01\rangle$ we obtain the classical bimatrix with the only
difference that the columns of the bimatrix are permuted. Another
case, in which the initial state $(|00\rangle +
|11\rangle)/\sqrt{2}$ is considered, could be interpreted that
players play with equal probability the classical game and the
game where both rows and columns were permuted. Of course, such
interpretation corresponds to outcomes given by the MW protocol
where the payoff pair is the same irrespective of choosing
$I\otimes I$ or $C\otimes C$ (see, for instance, the Battle of the
Sexes game studied in \cite{mw}).
\\ \noindent It turns out that this natural feature of the MW scheme is not
transferred
 to Iqbal and Toor's generalization. For technical convenience, let us number the
computational states from 0 to 2 instead of Iqbal and Toor's
numbering from 1 to 3. Definitions of unitary strategies $I$, $D$
and $C$ introduced in
 \cite{iqbalfail}, but with respect to numbering from 0 to 2 are as follows:
\begin{equation} \begin{matrix}
I|0\rangle = |0\rangle &~~~~ C|0\rangle = |2\rangle &~~~~
D|0\rangle =
|1\rangle\\
I|1\rangle = |1\rangle &~~~~ C|1\rangle = |1\rangle &~~~~
D|1\rangle =
|0\rangle\\
I|2\rangle = |2\rangle &~~~~ C|2\rangle = |0\rangle &~~~~
D|2\rangle = |2\rangle
\end{matrix}
\end{equation}
and they mean, respectively, the first, the second and the third
strategy for each player. Although their protocol yields the
classical $3 \times 3$ game if $|\psi_{\mathrm{in}}\rangle =
 |00\rangle$, it gives nonequivalent games for other basis states.
Putting, for example, $|\psi_{\mathrm{in}}\rangle = |01\rangle$,
the bimatrix of a $3 \times3$ game changes as follows:
\begin{equation} \label{equation1}
\begin{pmatrix} P_{00} & P_{01} & P_{02} \\ P_{10} & P_{11} &
P_{12}\\ P_{20} & P_{21} & P_{22}
\end{pmatrix} \xrightarrow{|\psi_{\mathrm{in}}\rangle =
|01\rangle}
 \begin{pmatrix} P_{01} & P_{00} & P_{01} \\ P_{11} &
P_{10} & P_{11}\\ P_{21} & P_{20} & P_{21}
\end{pmatrix},
\end{equation} where $P_{ij}:=(a_{ij}, b_{ij})$. The output game
is not equivalent (up to the order of the columns) to the input
one. Instead, we obtain the game in which the outcomes $P_{02}$,
$P_{12}$ and $P_{22}$ are no longer available. Obviously, such a
change affects a game significantly. However, in our opinion,
transformation (\ref{equation1}) should not be caused by another
basis state. According to theory of quantum correlations, superior
results can be created only by entangled states \cite{lewen}.
Therefore non-classical transformation (\ref{equation1}) obtained
simply by a separable basis state suggests some irregularity in
the Iqbal and Toor's approach. Obviously, the undesirable property
(\ref{equation1})
translates into any superpositions of states.\\
\noindent It is not difficult to note that the problem lies in the
definition of operators $D$ and $C$. Although the players are
provided with the identity operator $I$, the operators $C$ and $D$
act like the identity operator for states $|1\rangle$ and
$|2\rangle$, respectively. In consequence, player 2 is not able to
change her state from $|1\rangle$ to $|2\rangle$ which makes some
of the outcomes unavailable.
\section{The extension of the Marinatto-Weber scheme to any finite bimatrix
game} Consider an $n \times m$ bimatrix game
\begin{equation}\label{equation2}
\begin{pmatrix}
P_{00} & P_{01} & \cdots & P_{0,m-1}\\
P_{10} & P_{11} & \cdots & P_{1,m-1}\\
\vdots & \vdots & \ddots & \vdots\\
P_{n-1,0} & P_{n-1,1} & \cdots & P_{n-1,m-1}
\end{pmatrix},
\end{equation}
where $P_{ij} \in \mathbb{R}\times\mathbb{R}$ and define $l$
unitary operators $V_{k}$ for $k = 0,1,\dots,l-1$,  that act on
states of the computational basis $\{|0\rangle, |1\rangle, \dots
|l-1\rangle\}$ as follows:
\begin{equation}
 \begin{array}{l}
V_{0}|i\rangle = |i\rangle, \\ V_{1}|i\rangle = |i + 1 \bmod l\rangle, \\
~~~~~~~~~~~~~~~~\vdots  \\ V_{l-1}|i\rangle = |i + (l-1) \bmod l
\rangle.
\end{array}
\end{equation}
According to this extension of the MW scheme, for the game
(\ref{equation2}) one obtains the final state
\begin{equation}\label{equation4}
\rho_{\mathrm{fin}} = \sum^{n-1}_{i =
0}\sum^{m-1}_{j=0}p_{i}q_{j}V_{i}\otimes
V_{j}\rho_{\mathrm{in}}V^{\dag}_{i}\otimes V^{\dag}_{j},
\end{equation}
where $\{p_{i}\}$ and $\{q_{j}\}$ are probability distributions
over $\left\{V_{i}\right\}$ and $\left\{V_{j}\right\}$ and
$\rho_{\mathrm{in}}$ is the density operator for a state
$|\psi_{\mathrm{in}}\rangle \in \mathbb{C}^n\otimes \mathbb{C}^m$,
and the payoff operator
\begin{equation}
X = \sum^{n-1}_{i=0}\sum^{m-1}_{j=0}P_{ij}|ij\rangle \langle ij |.
\end{equation}
Then the average payoff pair is given by the formula
\begin{equation}\label{przedostatni}
E = \mathrm{tr}(X\rho_{\mathrm{fin}}).
\end{equation}
Let us determine the output bimatrix corresponding to the initial
state $|ij\rangle$ in order to prove the correctness of the scheme
(\ref{equation2})--(\ref{przedostatni}). Equations
(\ref{equation4})--(\ref{przedostatni}) imply the following
bimatrix:
\begin{equation}
\begin{pmatrix}\label{ostatni}
P_{ij} & \cdots & P_{i, j+m-1 \bmod m} \\
P_{i+1 \bmod n,j} & \cdots & P_{i+1 \bmod n, j+m-1 \bmod m}\\
\vdots & \ddots & \vdots\\
P_{i+n-1 \bmod n,j} & \cdots & P_{i+1 \bmod n, j+m-1 \bmod m}
\end{pmatrix}.
\end{equation}
As a result, we obtain the bimatrix which is equivalent to
(\ref{equation2}) from a game-theoretic point of view. The only
difference between them is that now rows and columns are numbered
from $i$ and $j$ instead of zero. In particular, in the problem
(\ref{equation1}) the bimatrix on the right hand side takes now
the following form:
\begin{equation}
\begin{pmatrix} P_{01} & P_{02} & P_{00} \\ P_{11} &
P_{12} & P_{10}\\ P_{21} & P_{22} & P_{20}
\end{pmatrix}.
\end{equation}
and it differs from the original one in the numbering order.
\section{Conclusion}
The idea of playing $3 \times 3$ games and the examination of
evolutionarily stable strategies in the language of quantum games
propounded by Iqbal and Toor in \cite{iqbalfail} and
\cite{iqbalfail2} show a great area of research into quantum
games. It suggests that the number of open problems in theory of
quantum games may be equal to the number of classical game theory
concepts that have not been studied yet in the quantum mechanics
environment.\\ \noindent We have refined Iqbal and Toor's
generalization to inherit the MW-scheme features.  Surprisingly,
the output game (\ref{ostatni}) coincides with Iqbal and Toor's
one \cite{iqbalfail} if the input game is their modified
Rock-Scissors-Paper game and the final state is $1/2(|01\rangle +
|10\rangle+ |02\rangle +|20\rangle)$. It follows from the fact
that the Rock-Scissors-Paper game is defined only by three
different numbers. In general, bimatrix (\ref{ostatni}) provides
us with a quite different output game than the one defined by
Iqbal and Toor. Therefore, we claim that our scheme scheme ought
to be used in the case of studying more complex games.

\section*{ACKNOWLEDGMENTS}
The author is very grateful to his supervisor Prof. J. Pykacz from
the Institute of Mathematics, University of Gda\'nsk, Poland for
his great help in putting this paper into its final form. The
project was supported by the Polish National Science Center under
the project number DEC-2011/03/N/ST1/02940.

\end{document}